\documentclass[%
 reprint,
 amsmath,amssymb,
 aps,
]{revtex4-2}

\usepackage{graphicx}
\usepackage{dcolumn}
\usepackage{bm}

\begin{document}

\preprint{APS/123-QED}


\author{Dennis Scheidt}
\author{Pedro A. Quinto Su}%
 \email{Pedro.quinto@nucleares.unam.mx}
\affiliation{%
 Instituto de Ciencias Nucleares, Universidad Nacional Aut\'onoma de M\'exico, \\ Apartado Postal 70-543, 04510, Cd. Mx., M\'exico
}%


\date{\today}
\title{Comparison between Hadamard and canonical bases for in-situ wavefront correction and the effect of ordering in compressive sensing}

\begin{abstract}
In this work we compare the Canonical and Hadamard bases for in-situ wavefront correction of a focused Gaussian beam using a spatial light modulator (SLM). The beam is perturbed with a transparent optical element (sparse) or a random scatterer (both prevent focusing at a single spot). The phase corrections are implemented with different basis sizes ($N=64, 256, 1024, 4096$) and the phase contribution of each basis element is measured with 3 step interferometry. The field is reconstructed from the complete $3N$ measurements and the correction is implemented by projecting the conjugate phase at the SLM. 
Our experiments show that in general, the Hadamard basis measurements yield better corrections because every element spans the relevant area of the SLM, reducing the noise in the interferograms. %
In contrast, the canonical basis has the fundamental limitation that the area of the elements is proportional to $1/N$, and requires dimensions that are compatible with the spatial period of the grating. 
In the case of the random scatterer, we were only able to get reasonable corrections with the Hadamard basis and the intensity of the corrected spot increased monotonically with $N$, which is consistent with fast random changes in phase over small spatial scales.
We also explore compressive sensing with the Hadamard basis and find that the minimum compression ratio needed to achieve corrections with similar quality to those that use the complete measurements depend on the basis ordering. The best results are reached in the case of the Hadamard-Walsh and cake cutting orderings. 
Surprisingly, in the case of the random scatterer we find that moderate compression ratios on the order of $10-20\%$ ($N=4096$) allow to recover focused spots, although as expected, the maximum intensities increase monotonically with the number of measurements due to the non sparsity of the signal.
\end{abstract}


\maketitle
\section{Introduction}
Most optical instruments and applications can be very sensitive to aberrations. However, programmable optical elements like deformable mirrors and 2 dimensional arrays like spatial light modulators (SLM) and digital micromirror arrays (DMD) have enabled corrections of arbitrary errors that can result in better focusing and enhancing light transmission through scattering materials \cite{Brousseau:07,Weiss:18}.

A very simple solution to correct aberrations with a spatial light modulator is the `in-situ' method that cancels the errors at the point where the beam is used (i.e. focused spot) demonstrated by \v Ci\v zmar and coworkers \cite{insitu1}. %
That approach uses an orthogonal canonical basis representation of the field at the surface of an SLM, where the basis is implemented in a square domain with an area $A$ that is divided into a grid of $N$ non-overlapping elements. Each element is represented by a unit vector in $N$ dimensional space $\hat e_i=\delta _{ij}$ ($i,j=1,...,N$). 
The amplitude and phase contribution of each square wavelet at the point of interest is measured with interferometry. In that way, it is possible to enhance the intensity of the focused beam by assigning each mode a phase that maximizes the interference, cancelling the aberrations and enhancing the intensity. 
This method has also been implemented with a DMD that has faster update rates \cite{Zupancic:16} to generate very accurate structured beams.

A fundamental limitation of the in-situ method with the canonical basis is that increasing the spatial resolution of the measurement by increasing the size of the basis $N$ decreases the area of the individual elements to $A/N$, reducing the amount of light diffracted by each element which can increase noise in the measurement. Furthermore, if the size becomes smaller than the grating period that is used to diffract the light, the diffraction efficiency decreases exacerbating the problem.

Other orthogonal bases (i.e. Hadamard) utilize overlapping modes that fill the relevant area $A$ of the SLM where the basis is defined \cite{SinglePixel}. For example, the $N$ orthogonal vectors of the Hadamard basis do not have zero entries but $\pm 1$ values; as a result the signal to noise ratio increases and is similar for all elements. 

Furthermore, the spatial overlap of the modes in the case of the Hadamard basis allows the use of compressive sensing (CS) \cite{reviewsinglepixel}, where the full signal can be reconstructed measuring only a subset $M$ of vectors in the basis ($M<N$). This can be done because most signals are sparse with only a few components contributing significantly. An important consideration in the implementation of compressive sensing is the ordering of the Hadamard basis  \cite{HadOrder,HadRussDoll,HadOrigami}, which can significantly decrease the number of measurements needed to reconstruct a signal.

Several groups have already implemented compressive sensing for imaging a complex field (phase and amplitude) \cite{SinglePixel}, but to our knowledge there are no studies comparing the original 'in-situ' experiment with a Hadamard basis to correct a focused beam, which is important for many applications like micromanipulation and microfabrication. 

In this work we demonstrate in-situ wavefront correction for two extreme cases: a sparse perturbation with a transparent optical element and a random scatterer. 
In general, we find that the signal to noise ratio degrades for the largest canonical basis (N=4096), while the case of the Hadamard basis is independent of $N$.
The case of the random scatterer cannot be corrected with the canonical basis, while the Hadamard correction improves monotonically with $N$. 
We also investigate the following orderings of the Hadamard basis for compressive sensing: Hadamard, Hadamard-Walsh, cake cutting and random. The best results are achieved with the Hadamard-Walsh and cake cutting orderings. 

This article is arranged in the following way:
The experimental setup is described in section 2, including the implementation of the bases at the SLM, the 3 step interferometry to recover the phase and amplitude contribution of each element and the description of the perturbing elements. The results of the full measurements ($3N$) of the bases are presented in section 3. Compressive sensing is described in section 4 and the results are in section 5 where wavefront corrections are implemented using subsets of the full measurements (between $2-90\%$) and different orderings of the Hadamard basis (Hadamard, Hadamard-Walsh, cake cutting, random). Section 6 contains the concluding remarks.

\section{Experiment}
This section contains details about the experimental setup, bases, encoding at the SLM, 3 step interferometry, field reconstruction and specific details about the perturbations and the realization of the experiment.

\subsection{Experimental Setup}
The experiment schematic (not to scale) is depicted in Figure 1a, where a CW laser beam ($\lambda = 1064\,$nm, $4.95\pm 0.01\,$mW) with linear polarization controlled by a half wave plate (HWP1) is reflected by a spatial light modulator (SLM - Hamamatsu: LCOS-SLM X10468, $20\, \mu$m pixel size) that imprints a 2d prism phase to the horizontal polarization component to separate it from the zero order. %
Then a polarizing beam splitter cube (PBS) separates the modulated horizontal and unmodulated vertical polarization that makes the reference beam which propagates through a small aperture (diameter of $\sim 1.5\,$mm) and then through a half wave plate (HWP2) projecting it into the horizontal polarization state. A lens ($f=40\,$cm) at a distance of $40\,$cm from the SLM focuses the beams into the camera (Thorlabs DCC1645C, $3.6\,\mu$m pixel size), that can acquire the transverse intensity profile of the uncorrected/corrected beams (with no reference) or the interferograms with are measured at a single pixel. 
We introduce an additional element to the modulated beam at an arbitrary position to perturb it preventing a good focus: a transparent element or a random scatterer (Fig 1a, D1 and D2). 

The experiment is similar to the original in-situ article \cite{insitu1} and the main difference is the interferometer. \v Ci\v zmar and coworkers extracted the reference beam from a single basis  element located at the center of the grid, while our reference emerges from a small section (controlled with the iris) of the unmodulated beam which is equivalent. Both experiments perform the measurements with a CCD camera, which allows to quickly measure the transverse profiles of the beams.

\begin{figure} 
\centering    
\includegraphics[width = 0.4\textwidth]{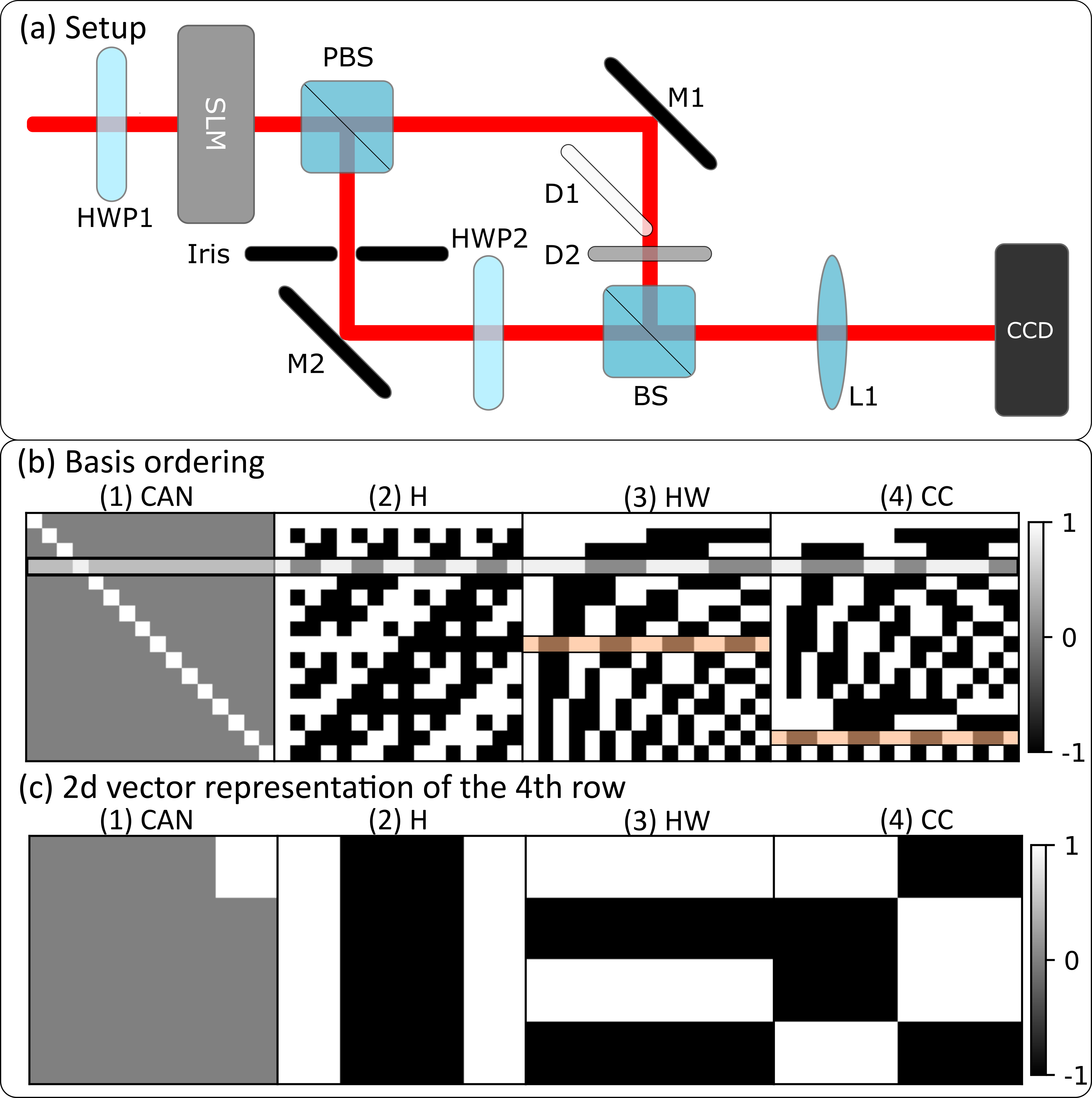}
\caption{Experimental setup. (a) Experimental setup. 
(b) Measurement bases in their matrix representation $\Phi$ ($16\times 16$ elements): (1) CAN: Canonical (2) H:  Natural Hadamard (3) HW: Hadamard-Walsh (4) CC: cake-cutting. The 4th row is highlighted in all $\Phi$, the orange highlighted components in HW and CC show the position of the 4th element of H showing the different ordering. (c) 2d vector representation of the 4th 1d-measurement vector. }
\end{figure}

\subsection{Bases and 2d representation at the SLM}
We consider the canonical and the Hadamard bases. The bases are represented by matrices with $N\times N$ elements, and each vector (row in the matrix) is represented in the SLM as a square grid with $\sqrt{N}\times \sqrt{N}$ that is resized to the size of the screen area that encodes the basis at the SLM ($512\times 512\,$px). 

\subsubsection{Canonical basis}
The canonical basis is represented by an $N\times N$ identity matrix $\Phi _{CAN}=I_N$, where the orthogonal vectors are the rows or columns.
The first frame of Fig. 1b shows $\Phi _{CAN}$ for $N=16$ with the fourth element highlighted, the encoding of that element at the SLM screen is shown in the first row of Fig. 1c ($\sqrt{N}\times \sqrt{N}$ array). 

\subsubsection{Hadamard basis}
The Hadamard basis $\Phi _H$ ($N\times N$ matrix) is obtained from the Hadamard matrix, which is made of $\pm 1$ entries.
The Hadamard basis can be ordered in different ways (natural, Hadamard-Walsh, cake-cutting, random) which are relevant to compressive sensing (sections 4 and 5). 

In this article we refer to the natural ordering of the basis as Hadamard (H) and the second row of Fig. 1b shows $\phi _H$ for $N=16$ and in the next row (Fig. 1c) the encoding of the fourth element at the SLM.

The Walsh ordered Hadamard basis (HW)  $\Phi _{HW}$ orders the Walsh functions according to the increasing number of zero crossings, acting as a low pass filter \cite{HadOrder}. The case for $N=16$ of $\Phi _{HW}$ is in the third column of Fig. 2b, and the encoding of the fourth element at the SLM is depicted in the third row of figure 1c as an $\sqrt{N}\times \sqrt{N}$ array. 
 
The cake-cutting ordering ($\Phi_{CC}$) is calculated by displaying the $1 \times N$ row-vectors of the Hadamard basis $\Phi_{H}$ in the $\sqrt{N} \times \sqrt{N}$ representation that is projected onto the SLM. These row-vectors are then ordered according to the number of connected "$+1$"- entries in their 2d representation \cite{CakeCutting}. The encoding at the SLM of the fourth element of $\phi _{CC}$ is shown in the last row of Fig. 1c. 
To illustrate the different ordering between HW and CC one of the elements is highlighted in orange, showing the change in ordering.

The random ordering (HRAN) of the Hadamard basis ($\Phi _{HRAN}$) is a random permutation of the vectors (rows) of $\Phi _H$. 

\subsection{Encoding at the SLM}
The Gaussian beam is reflected at the SLM and diffracted with a linear prism phase (not shown, spatial period$\sim 20\,$px) in order to separate it from the unmodulated zero order. In order to imprint the basis elements (in the 2d representation) into the SLM, we consider a constant area $A$ so that all the different basis sizes $N$ area resized to fill it (i.e. $512\times 512$ px).

In the case of the canonical basis, the prism phase is simply multiplied by the 2d representation of each vector element that does not vanish at a square. In that way, the light is only diffracted by the individual square wavelets.

In the case of the Hadamard basis where each element has $+1$ and $-1$ entries, normally each element has to be sampled twice to account for the different signs when performing intensity measurements like those involving photographs or images. Here, in order to avoid double sampling \cite{reviewsinglepixel} we use the approach of \cite{ota} where the $+1$ basis elements have a phase difference of $\pi$ compared with the $-1$. Hence, a phase of $\pi$ is added to the prism at the locations that correspond to a $-1$ value in the basis element. In that way, the subtraction is achieved in a single step.

The field is represented by the complex vector $x=|x| e^{i \phi _x}$ ($1\times N$ elements) that has to be measured and is also encoded as a $\sqrt{N}\times \sqrt{N}$ array at the SLM. In order to compare the effect that the basis $\Phi$ has on the experiment, $x$ is sampled with the canonical and the Hadamard representation. 
%
\begin{figure}  
\centering
\includegraphics[width = 0.45 \textwidth]{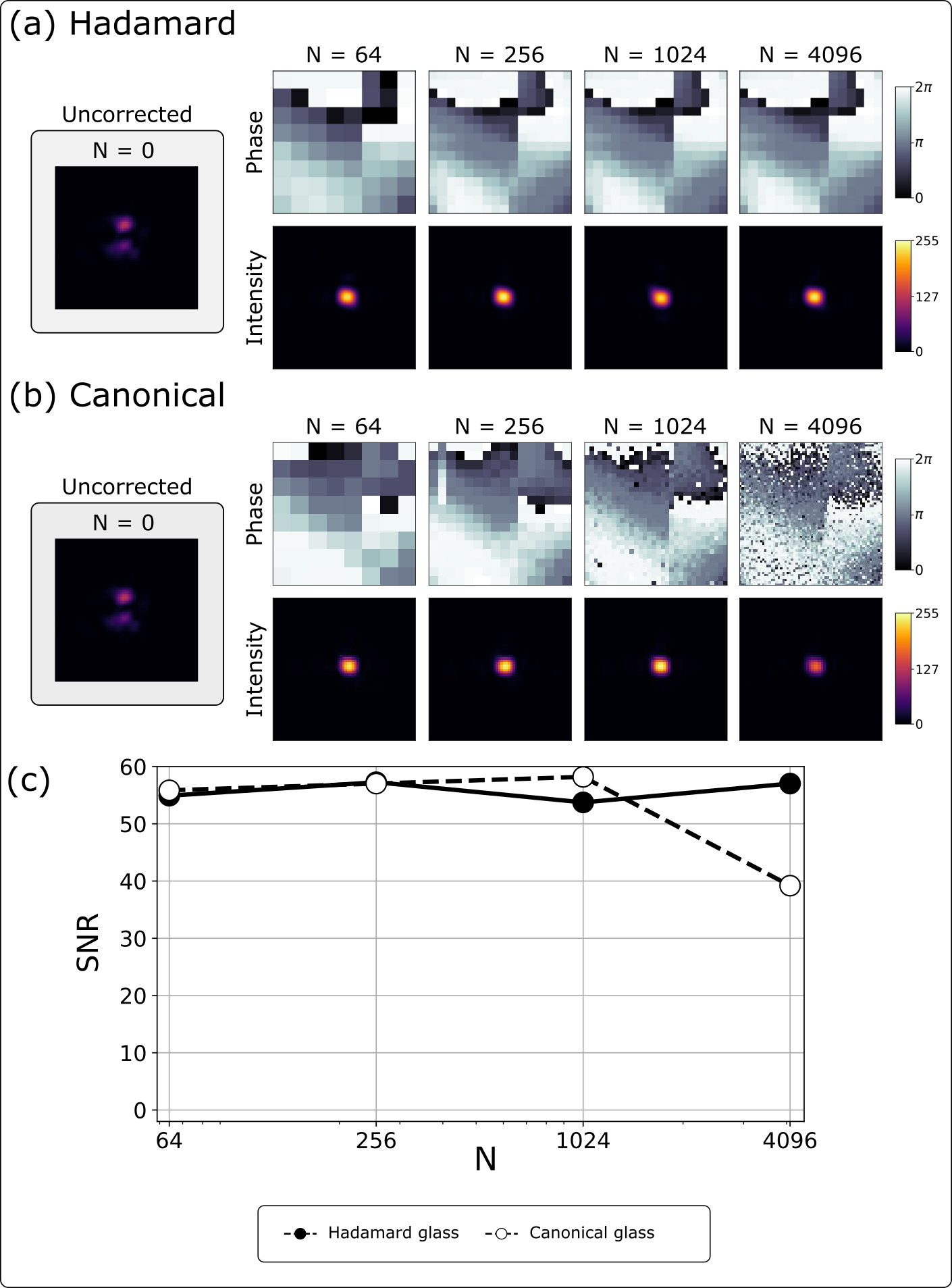}
\centering
\caption{Reconstruction results for the full bases in glass: Reconstructed phase, and corrected spot for $N$ between $64$ and $4096$ with the Hadamard basis (a) and the canonical basis (b). 
(c) Signal to noise ratio (SNR) of the corrected spots as a function of $N$ for the Hadamard (filled circles) and canonical (hollow circles) basis.}
\end{figure}
%
%
\subsection{3 step interferometry}
In order to reconstruct the phase and amplitude of the perturbed field $x$, the interference between the reference beam and each individual mode of a basis $\Phi$ is measured using three-step interferometry by adding constant phase shifts $\delta _m =0$, $2 \pi /3$ and $4 \pi /3$ (m=1-3).
The interference between the full field $x$ and the reference beam for a given phase shift is represented by $x_m$.
In this way the measured interference is (in arbitrary units) $x_m = I_r +I_x+2\sqrt{I_r I_x} \cos{(\Delta \phi +\delta _m)}$, where $I_r$ is the intensity of the reference beam, $I_x \propto |x|^2$ the intensity of the field at the SLM and $\Delta \phi=\phi _x -\phi _r$ with $\phi _x$ and $\phi _r$ the phase of $x$ and the reference, respectively. The reference beam contributes with a constant amplitude and phase at the point of detection. The measured interference between the different elements of the basis $\Phi $ and the reference for a given phase shift are $I_{i,m}$ ($i=1,...,N$).

\subsection{Field reconstruction}
The full set of measurements ($3N$) is grouped into 3 vectors $I_m$ that contain the individual measurements $I_{i,m}$ for a given phase shift $\delta _m$. In this way 
\begin{equation}
\Phi x_m = I_m, 
\end{equation}
the vectors $x_m$ are recovered by solving the system of equations ($x_m = \Phi ^{-1} I_m$) and the full complex field $x$ is recovered (up to a constant amplitude and phase) with \cite{Zupancic:16}:
\begin{equation}
    x = -\frac{1}{3}(x_{2} + x_{3} - 2x_{1}) + \frac{i}{\sqrt{3}} (x_{2} -x_{3})
\end{equation}
The (effect of the perturbations) aberrations are corrected by projecting the conjugate of $\mathrm{arg}(x)$ at the SLM. 
%
%
\subsection{Perturbations and realization of the experiment}
Two different perturbations are added to the modulated beam: transparent element (glass slide) which is representative of regular aberrations or a random scatterer made of Parafilm (Fig. 1a). Notice that the position of these elements (D1 and D2 respectively) is arbitrary (just like in the original in-situ experiment \cite{insitu1}). 

The glass slide cuts the beam at an angle through the center, creating a phase discontinuity resulting in a couple of focused spots (insets of Figure 2a,b). The exposure time for the focused spots with no reference (uncorrected and corrected) is $0.1\,$ms, while the exposures for the interferometric measurements are $1\,$ms for the Hadamard basis (independent of $N$) and on the order of tens of milliseconds for the canonical basis, with the longest exposure for the largest $N$ (we set 65\,ms as the global maximum to reduce the time to perform all the experiments).

The random scatterer is made of Parafilm (element D2, Fig. 1a) which has been used by several groups  \cite{Skarsoulis:21,Kanngiesser:19,Kanngiesser:Nature,Boniface:19}. The resulting focused beam (with no reference) is a very dim speckle pattern. The exposure time for the intensity measurements (uncorrected and corrected) is $0.65\,$ms, while the exposures for the interferometric measurements are $20\,$ms for Hadamard and between $30-65\,$ms for the canonical basis. Notice that for the random scatterer the exposure times have increased because multiple scattering drastically reduces the amount of light that is transmitted. 

For a given perturbation the experiment is conducted in the following way: First, HWP1 is rotated to prepare the beam in horizontal polarization, so there is no reference. Then the uncorrected beam is photographed. Next, HWP1 prepares the beam in a mixed polarization (Horizontal $+$ Vertical) state in order to have a reference and to maximize the visibility of the interferogram for the first basis size $N=64$ while the intensity of each of the $3N$ interferograms is captured at a single pixel in the camera. The field is reconstructed with the $3N$ measurements and the conjugate phase is projected at the SLM, HWP1 prepares the horizontal polarization and the corrected beam is photographed. In the case of the Hadamard basis, the measurements are reordered according to H, HW, CC and HRAN. For each ordering, the field is reconstructed with compressive sensing for a given subset of $N$ (between 2 and $90\%$) and the corrected beam is photographed. Once the measurements for Hadamard and canonical bases are completed, the process is repeated for the next $N$.

%
\subsection{Signal to noise ratio}
In order to quantify the quality of the corrected beam, it is compared to the uncorrected one with the signal to noise ratio (SNR).
The SNR is defined as the ratio between the maximum intensity of the corrected spot ($\text{Max}(I_{corr})$) and the mean of the uncorrected beam $\langle I_{uncorr}\rangle$ \cite{Vellekoop:07,T_Mat1}:
\begin{equation}
    SNR=\frac{\text{Max}(I_{corr})}{<I_{uncorr}>}.
\end{equation}
\begin{figure}  
\centering
\includegraphics[width = 0.45 \textwidth]{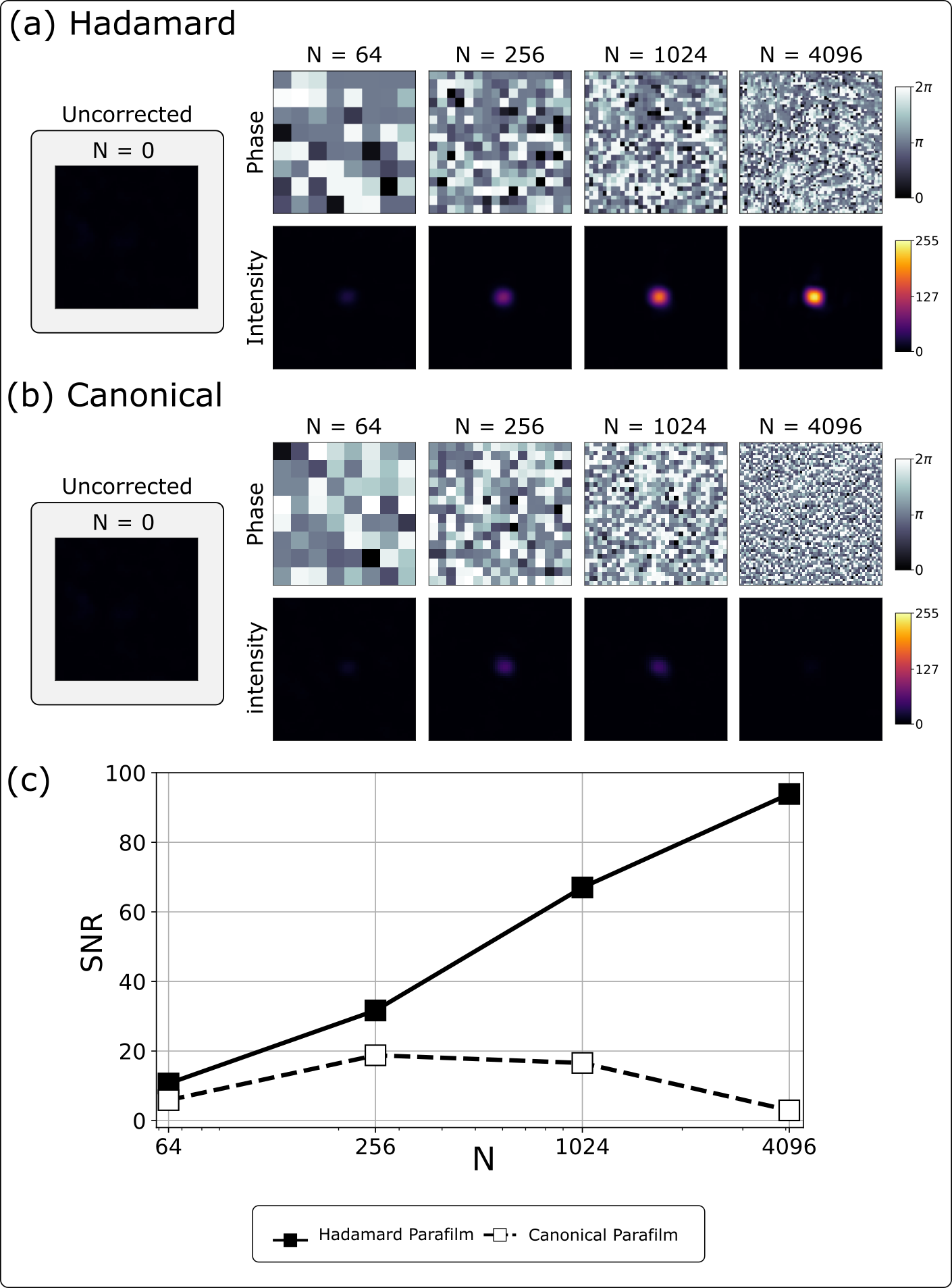}
\centering
\caption{Reconstruction results for the full bases in Parafilm: Reconstructed phase, and corrected spot for $N=64$ and $4096$ with the Hadamard basis (a) and the canonical basis (b). 
(c) SNR of the corrected spots as a function of $N$ for the Hadamard (filled squares) and canonical (hollow squares) basis..}
\end{figure}
%

\section{Full measurements and Discussion}
This section describes the results for both transparent optical element and random scatterer consider the full $3N$ measurements.

\subsection{Glass perturbation (transparent element)}
The results for the transparent element considering the full measurements are in Fig. 2. 
The reconstructed phase and corrected spots for the Hadamard basis ($N=64$ to $4096$) and are in Fig. 2a, while the inset to the left shows the uncorrected spot.
The results for the canonical basis are in Fig. 2b. We observe that the phase reconstructions have less noise in the case of the Hadamard basis, while the phases reconstructed with the canonical phase have a sudden increase in noise at $N=4096$.

The quantitative comparison for the corrected spots via a SNR as a function of $N$ is in Fig. 3C. The filled circles represent the SNR with the Hadamard basis while the hollow circles are those with the canonical basis. 
We observe that in the case of the Hadamard basis the SNR is fairly constant with a mean of 56 and a variation of $4\%$ when considering all $N$. In the case of the canonical basis, the trend is similar up to $N=1024$, however at $N=4096$ there is a significant decrease in the SNR to a value of $39$. This is because the size of the square wavelets is $8$ pixels at the SLM which is not compatible with the prism phase that has a larger spatial period ($\sim 20\,$px). As a result the diffraction efficiency decreases along increasing the measurement noise, showing the fundamental limit of the canonical basis.

\subsection{Random scatterer}
The reconstructed phases and measured spots when the beam is perturbed by the random scatterer are in Fig. 3.
In contrast with the glass perturbation, the dispersive Parafilm randomizes the phase of the transmitted light. The reconstructed phases and measured corrected focused spots with the Hadamard and canonical bases are in Fig. 3(a) and (b) respectively. The reconstructed phases confirm the randomizing effect that the Parafilm has on the beam. The uncorrected spots are shown in insets to the left. 
We observe that the intensity of the corrected spots increase monotonically with $N$ in the case of the Hadamard measurements, while those of the canonical basis are almost indistinguishable from the uncorrected spot.

The SNR data as a function of $N$ is in Fig. 3c. The measurements with the Hadamard basis (filled squares) show the same trend as the photographs in Fig. 3a, where the maximum SNR is reached at $N=4096$ with a value $\sim 94$. 
The canonical basis has a very poor performance (hollow squares) at all $N$ with a maximum SNR of 20 at $N=256$. Part of the reason is that the maximum exposure time of 65\,ms is not enough to acquire a good interferogram of each small square wavelet as the light is multiply scattered by the Parafilm. Again, the smallest SNR (value of 3) for the canonical basis is reached at $N=4096$ where the low light signal problem is exacerbated by the decrease in diffraction efficiency. 

Notice that the SNR reached for Parafilm are larger than for the glass element, this is because the average intensity of the uncorrected glass perturbation is larger than that of uncorrected Parafilm. 
%
\begin{figure}  
\centering
\includegraphics[width = 0.45\textwidth]{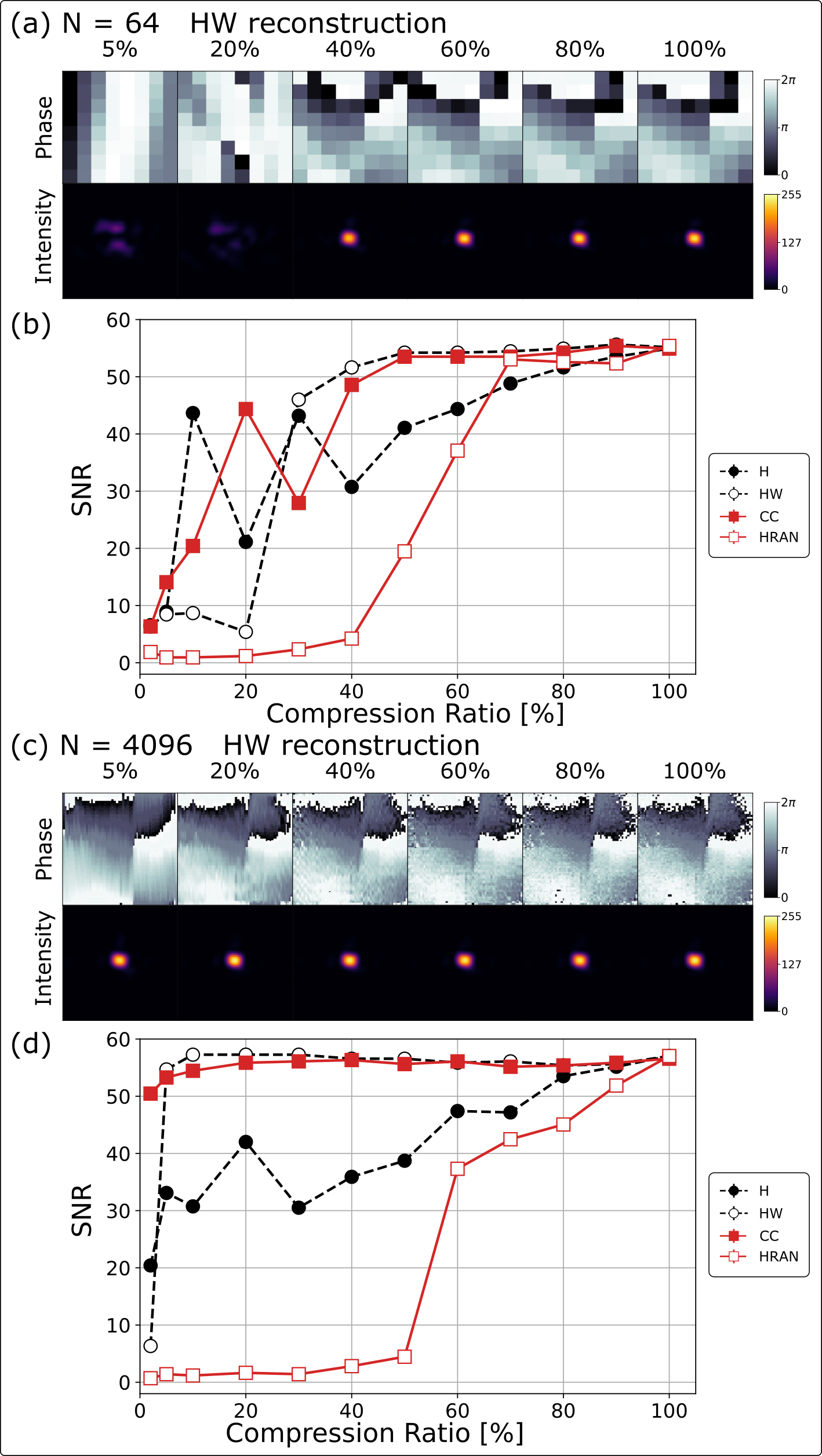}
\caption{a) Reconstructed phase and corrected spots for glass with the HW ordering with different $cr$ ($N=64$). b) SNR of the corrected spots as a function of $cr$ for all the orderings in the case of $N=64$. c) Reconstructed phase and corrected spots for glass with the HW ordering with different $cr$ ($N=4096$). d) SNR of the corrected spots as a function of $cr$ for all the orderings with $N=4096$.}
\end{figure}
\section{Compressive Sensing}
The compressive sensing (CS) approach to solving eq. (1) consists of using a smaller number $M$ of elements ($M<N$) where $M$ depends on the  sparsity of the representation for a given dataset with $K$ sparse entries: $M \approx K \log(N/K) $\cite{databook}.
The reduced basis is represented by $\Phi '$ which is an $N\times M$ matrix. The corresponding intensity measurement vectors ($1\times M$) are $I '_m$ with $\Phi ' x_m=I'_m$.
Most optical signals are sparse in frequency space, so we use the discrete cosine transform (dct) which is a reasonable sparse basis \cite{databook}. We express the field $x$ in this basis: $\mathrm{dct}(x_m) =s_m$ or $x_m =\Psi s_m$ with $\Psi =\mathrm{idct}(I_N)$ ($I_N$ is the $N\times N$ identity matrix and idct the inverse discrete cosine transform). The relation between $s_m$ and the subset of measurements $I'_m$ is $\Phi ' \Psi s_m =I'_m$, which can be rewritten as $\Theta s_m = I' _m$ with $\Theta = \Phi ' \Psi$ the sensing matrix.

\noindent The problem is stated as a convex L1 minimization:
\begin{align}
    \text{min}\vert\vert s_m \vert\vert_1 \ \ \text{such that} \ \ I'_m = \Theta s_m \label{eq:CS}
\end{align}
\noindent This problem is solved using SPGL1 basis pursuit algorithm with its Matlab implementation \cite{BergFriedlander:2008, spgl1site}. 
The $x_m$ are recovered with $x_m=\Psi s_m$ and finally the complex field $x$ is extracted with eq. (2).
\section{Compressive sensing results and Discussion.}
The results using different compression ratios $cr=M/N$ considering different orderings of the Hadamard basis are presented in Figs. 4 and 5 for the transparent and random element respectively. 
In the SNR plots, the abbreviations and symbols used for the different orderings are: H for Hadamard (filled black circles), HW for Hadamard Walsh (hollow black circles), CC for cake cutting (filled red squares) and HRAN for random ordering (hollow red squares).
%

\begin{figure}  
\centering
\includegraphics[width = 0.45\textwidth]{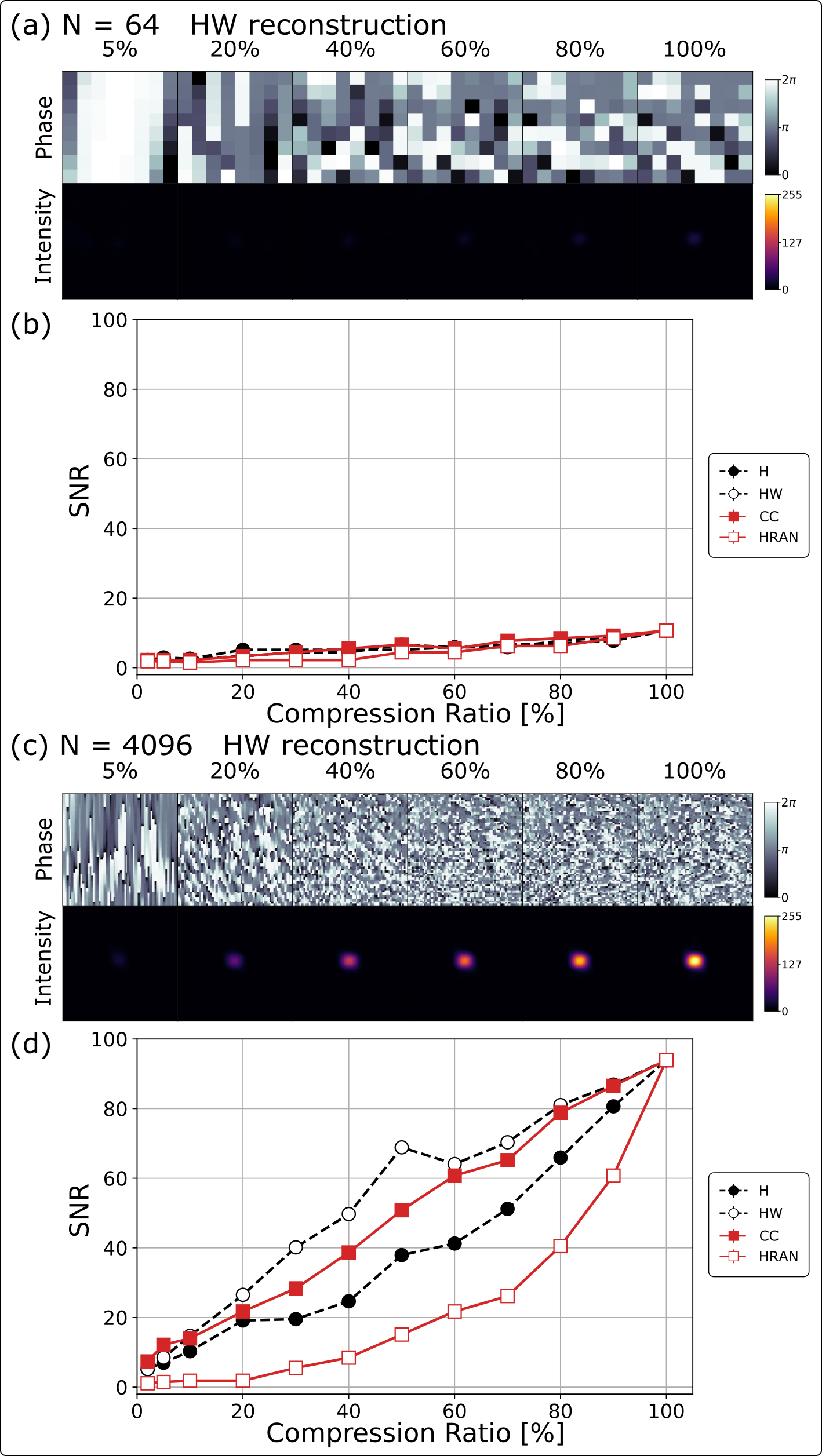}
\caption{a) Reconstructed phase and corrected spots for Parafilm with the HW ordering with different $cr$ ($N=64$). b) SNR of the corrected spots as a function of $cr$ for all the orderings in the case of $N=64$. c) Reconstructed phase and corrected spots with the HW ordering with different $cr$ in the case of $N=4096$. d) SNR as a function of $cr$ for all the orderings with $N=4096$.
}
\end{figure}

\subsection{CS Transparent element}
Figure 4a shows the phase reconstructions and corrected spots for the glass perturbation in the case of $N=64$ with the cake-cutting ordering with different $cr$. The SNR data considering all the orderings is plotted in Fig. 4b for $N=64$. As expected we observe that increasing $cr$ results in an enhanced SNR until it reaches the value of the full measurement, which for the transparent element is $\sim 56$ (Fig. 2c). SNR on the order $\geq 48$ are reached at $cr=40\%$ in the case of the HW and CC orderings, with SNR at the maximum value of $\sim 56$ at $cr=50\%$ and above. 

In the case of $N=256$, both the HW and CC orderings reach SNR of 52 at $cr=20\%$, while for $N=1024$ those orderings reach SNR above 45 at $cr=10\%$. 

Figure 4c shows the reconstructed phases and corrected spots for the largest basis with $N=4096$ elements for selected $cr$ for the HW ordering. The SNR data as a function of $cr$ for all orderings is plotted in Fig. 4d. Both, HW and CC orderings reach SNR above 53 at $cr=5\%$, while the H and HRAN orderings only reach similar values for $cr\geq 80\%$. 

To summarize the CS results for the transparent element, the best performing orderings are HW and CC, where CC converges at a slightly faster rate to the maximum SNR value. The H and HRAN orderings have a very slow convergence as a function of $cr$. We also observe that for HW and CC orderings the $cr$ at which SNR reaches values on the order of 45 decreases with increasing $N$. This means that for perturbations that are sparse we can increase the spatial resolution in the basis space and reconstruct the signal with a small subset of N. In the case of $N=4096$, $5\%$ of the full measurements ($3N$) is $614$ which is smaller than the full measurements ($3N=768$) for $N=256$. 

\subsection{CS Random scatterer}
Figure 5a shows phase reconstructions for the Parafilm scatterer and corrected spots at selected $cr$ for $N=64$ and the HW ordering. The images of the corrected spots are very similar to the uncorrected one (Fig. 3 a-b). The reason is that measurements with that spatial resolution yield poor results as the random scatterer induces very fast changes in phase at small spatial scales.
The data considering SNR as a function of $cr$ is in Fig. 5b, which shows that all the orderings yield similar results. The maximum SNR is that of Fig. 3c for the Hadamard basis (at $N=64$) which is smaller than 12. 
A similar trend is observed for $N=256$ (not shown) where H, HW and CC are slightly better than HRAN and all converge to the SNR $\sim 32$ for $cr=100\%$. The case for $N=1024$ has HW and CC with better SNR than H and HRAN for $cr <80\%$ and the maximum SNR is $\sim 65$. 

Figure 5c has the reconstructed phase and corrected spots for the HW ordering at $N=4096$. Interestingly, the corrected beams already show a focused spot at $cr$ between $10-20\%$, which the peak intensity increasing with $cr$.
The SNR as a function of $cr$ is plotted in Fig. 5d for all the orderings.
Again we observe a monotonic increase in SNR with $cr$ and the best performing orderings are HW and CC. However HW is better than CC for $cr\leq 80\%$ and HW reaches SNR $sim 70$ at $cr=50\%$. 

In the case of the random scatterer we observe a very different trend in the SNR data compared with the transparent optical element (sparse), there are no sudden jumps in SNR that approach those of the complete $3N$ measurement ($cr=100\%$), but monotonic increases for all the different orderings. 
Nevertheless, the Walsh and cake cutting orderings still outperform significantly the Hadamard and random. especially between $cr=30$ and $80\%$.

\section{Conclusion}
We compared measuring the field of a perturbed beam using the Hadamard and canonical bases. The perturbations we considered a transparent element and a random scatterer. In general, the use of the Hadamard basis to sample the field increases the accuracy of the reconstructed phase compared to the canonical basis and thus enables a significant improvement of the correction through highly scattering media. 

The results of the transparent element should be representative of other aberrations that have sparse representations like those described by the Zernike polynomials \cite{ZernnikeCS}. The compressive sensing experiments show that the ordering of the basis has a very large impact on the convergence of the SNR for a given $cr$. The best results for the compressive correction through glass are reached for HW and CC orderings. The plot of SNR as a function of $cr$ have the expected trends, with a threshold $cr$ where the SNR jumps to a value close to that of the full measurement. The value of $cr$ where that happens decreases with $N$ reaching a value of $5\%$ for $N=4096$.

The random scatterer represents the extreme case where there is no sparse representation and requires a large basis that can resolve the fast phase changes at small spatial scales. For these cases there is an advantage in using the full area where the basis is defined (like in the Hadamard basis), in contrast with small non overlapping sections of the canonical basis. Furthermore, modest compression ratios can already have a large impact on the SNR, so it is better to subsample a large basis than complete measurements with a small one that does not have the required spatial resolution.
In the case of the random scatterer the best results were achieved by the Hadamard-Walsh ordering with $N=4096$, where a reasonable correction was reached at a $cr$ of $50\%$ and a focused spot can be observed at $cr=10-20\%$.


\begin{thebibliography}{21}%
\makeatletter
\providecommand \@ifxundefined [1]{%
 \@ifx{#1\undefined}
}%
\providecommand \@ifnum [1]{%
 \ifnum #1\expandafter \@firstoftwo
 \else \expandafter \@secondoftwo
 \fi
}%
\providecommand \@ifx [1]{%
 \ifx #1\expandafter \@firstoftwo
 \else \expandafter \@secondoftwo
 \fi
}%
\providecommand \natexlab [1]{#1}%
\providecommand \enquote  [1]{``#1''}%
\providecommand \bibnamefont  [1]{#1}%
\providecommand \bibfnamefont [1]{#1}%
\providecommand \citenamefont [1]{#1}%
\providecommand \href@noop [0]{\@secondoftwo}%
\providecommand \href [0]{\begingroup \@sanitize@url \@href}%
\providecommand \@href[1]{\@@startlink{#1}\@@href}%
\providecommand \@@href[1]{\endgroup#1\@@endlink}%
\providecommand \@sanitize@url [0]{\catcode `\\12\catcode `\$12\catcode
  `\&12\catcode `\#12\catcode `\^12\catcode `\_12\catcode `\%12\relax}%
\providecommand \@@startlink[1]{}%
\providecommand \@@endlink[0]{}%
\providecommand \url  [0]{\begingroup\@sanitize@url \@url }%
\providecommand \@url [1]{\endgroup\@href {#1}{\urlprefix }}%
\providecommand \urlprefix  [0]{URL }%
\providecommand \Eprint [0]{\href }%
\providecommand \doibase [0]{https://doi.org/}%
\providecommand \selectlanguage [0]{\@gobble}%
\providecommand \bibinfo  [0]{\@secondoftwo}%
\providecommand \bibfield  [0]{\@secondoftwo}%
\providecommand \translation [1]{[#1]}%
\providecommand \BibitemOpen [0]{}%
\providecommand \bibitemStop [0]{}%
\providecommand \bibitemNoStop [0]{.\EOS\space}%
\providecommand \EOS [0]{\spacefactor3000\relax}%
\providecommand \BibitemShut  [1]{\csname bibitem#1\endcsname}%
\let\auto@bib@innerbib\@empty
\bibitem [{\citenamefont {Brousseau}\ \emph {et~al.}(2007)\citenamefont
  {Brousseau}, \citenamefont {Borra},\ and\ \citenamefont
  {Thibault}}]{Brousseau:07}%
  \BibitemOpen
  \bibfield  {author} {\bibinfo {author} {\bibfnamefont {D.}~\bibnamefont
  {Brousseau}}, \bibinfo {author} {\bibfnamefont {E.~F.}\ \bibnamefont
  {Borra}},\ and\ \bibinfo {author} {\bibfnamefont {S.}~\bibnamefont
  {Thibault}},\ }\bibfield  {title} {\bibinfo {title} {Wavefront correction
  with a 37-actuator ferrofluid deformable mirror},\ }\href
  {https://doi.org/10.1364/OE.15.018190} {\bibfield  {journal} {\bibinfo
  {journal} {Opt. Express}\ }\textbf {\bibinfo {volume} {15}},\ \bibinfo
  {pages} {18190} (\bibinfo {year} {2007})}\BibitemShut {NoStop}%
\bibitem [{\citenamefont {Weiss}\ and\ \citenamefont {Katz}(2018)}]{Weiss:18}%
  \BibitemOpen
  \bibfield  {author} {\bibinfo {author} {\bibfnamefont {U.}~\bibnamefont
  {Weiss}}\ and\ \bibinfo {author} {\bibfnamefont {O.}~\bibnamefont {Katz}},\
  }\bibfield  {title} {\bibinfo {title} {Two-photon lensless micro-endoscopy
  with in-situ wavefront correction},\ }\href
  {https://doi.org/10.1364/OE.26.028808} {\bibfield  {journal} {\bibinfo
  {journal} {Opt. Express}\ }\textbf {\bibinfo {volume} {26}},\ \bibinfo
  {pages} {28808} (\bibinfo {year} {2018})}\BibitemShut {NoStop}%
\bibitem [{\citenamefont {Cizmar}\ \emph {et~al.}(2010)\citenamefont {Cizmar},
  \citenamefont {Mazilu},\ and\ \citenamefont {Dholakia}}]{insitu1}%
  \BibitemOpen
  \bibfield  {author} {\bibinfo {author} {\bibfnamefont {T.}~\bibnamefont
  {Cizmar}}, \bibinfo {author} {\bibfnamefont {M.}~\bibnamefont {Mazilu}},\
  and\ \bibinfo {author} {\bibfnamefont {K.}~\bibnamefont {Dholakia}},\
  }\bibfield  {title} {\bibinfo {title} {In situ wavefront correction and its
  application to micromanipulation},\ }\href@noop {} {\bibfield  {journal}
  {\bibinfo  {journal} {Nature Photonics}\ }\textbf {\bibinfo {volume} {4}},\
  \bibinfo {pages} {388} (\bibinfo {year} {2010})}\BibitemShut {NoStop}%
\bibitem [{\citenamefont {Zupancic}\ \emph {et~al.}(2016)\citenamefont
  {Zupancic}, \citenamefont {Preiss}, \citenamefont {Ma}, \citenamefont
  {Lukin}, \citenamefont {Tai}, \citenamefont {Rispoli}, \citenamefont
  {Islam},\ and\ \citenamefont {Greiner}}]{Zupancic:16}%
  \BibitemOpen
  \bibfield  {author} {\bibinfo {author} {\bibfnamefont {P.}~\bibnamefont
  {Zupancic}}, \bibinfo {author} {\bibfnamefont {P.~M.}\ \bibnamefont
  {Preiss}}, \bibinfo {author} {\bibfnamefont {R.}~\bibnamefont {Ma}}, \bibinfo
  {author} {\bibfnamefont {A.}~\bibnamefont {Lukin}}, \bibinfo {author}
  {\bibfnamefont {M.~E.}\ \bibnamefont {Tai}}, \bibinfo {author} {\bibfnamefont
  {M.}~\bibnamefont {Rispoli}}, \bibinfo {author} {\bibfnamefont
  {R.}~\bibnamefont {Islam}},\ and\ \bibinfo {author} {\bibfnamefont
  {M.}~\bibnamefont {Greiner}},\ }\bibfield  {title} {\bibinfo {title}
  {Ultra-precise holographic beam shaping for microscopic quantum control},\
  }\href {https://doi.org/10.1364/OE.24.013881} {\bibfield  {journal} {\bibinfo
   {journal} {Opt. Express}\ }\textbf {\bibinfo {volume} {24}},\ \bibinfo
  {pages} {13881} (\bibinfo {year} {2016})}\BibitemShut {NoStop}%
\bibitem [{\citenamefont {Liu}\ \emph {et~al.}(2019)\citenamefont {Liu},
  \citenamefont {Zhao}, \citenamefont {Zhang}, \citenamefont {Gao},\ and\
  \citenamefont {Li}}]{SinglePixel}%
  \BibitemOpen
  \bibfield  {author} {\bibinfo {author} {\bibfnamefont {R.}~\bibnamefont
  {Liu}}, \bibinfo {author} {\bibfnamefont {S.}~\bibnamefont {Zhao}}, \bibinfo
  {author} {\bibfnamefont {P.}~\bibnamefont {Zhang}}, \bibinfo {author}
  {\bibfnamefont {H.}~\bibnamefont {Gao}},\ and\ \bibinfo {author}
  {\bibfnamefont {F.}~\bibnamefont {Li}},\ }\bibfield  {title} {\bibinfo
  {title} {Complex wavefront reconstruction with single-pixel detector},\
  }\href@noop {} {\bibfield  {journal} {\bibinfo  {journal} {Applied Physics
  Letters}\ }\textbf {\bibinfo {volume} {114}},\ \bibinfo {pages} {161901}
  (\bibinfo {year} {2019})}\BibitemShut {NoStop}%
\bibitem [{\citenamefont {Gibson}\ \emph {et~al.}(2020)\citenamefont {Gibson},
  \citenamefont {Johnson},\ and\ \citenamefont {Padgett}}]{reviewsinglepixel}%
  \BibitemOpen
  \bibfield  {author} {\bibinfo {author} {\bibfnamefont {G.~M.}\ \bibnamefont
  {Gibson}}, \bibinfo {author} {\bibfnamefont {S.~D.}\ \bibnamefont
  {Johnson}},\ and\ \bibinfo {author} {\bibfnamefont {M.~J.}\ \bibnamefont
  {Padgett}},\ }\bibfield  {title} {\bibinfo {title} {Single-pixel imaging 12
  years on: a review},\ }\href {https://doi.org/10.1364/OE.403195} {\bibfield
  {journal} {\bibinfo  {journal} {Opt. Exp.}\ }\textbf {\bibinfo {volume}
  {28}},\ \bibinfo {pages} {28190} (\bibinfo {year} {2020})}\BibitemShut
  {NoStop}%
\bibitem [{\citenamefont {Vaz}\ \emph {et~al.}(2020)\citenamefont {Vaz},
  \citenamefont {Amaral}, \citenamefont {Ferreira}, \citenamefont {Morgado},\
  and\ \citenamefont {{a}o Cardoso}}]{HadOrder}%
  \BibitemOpen
  \bibfield  {author} {\bibinfo {author} {\bibfnamefont {P.~G.}\ \bibnamefont
  {Vaz}}, \bibinfo {author} {\bibfnamefont {D.}~\bibnamefont {Amaral}},
  \bibinfo {author} {\bibfnamefont {L.~F.~R.}\ \bibnamefont {Ferreira}},
  \bibinfo {author} {\bibfnamefont {M.}~\bibnamefont {Morgado}},\ and\ \bibinfo
  {author} {\bibfnamefont {J.}~\bibnamefont {{a}o Cardoso}},\ }\bibfield
  {title} {\bibinfo {title} {Image quality of compressive single-pixel imaging
  using different hadamard orderings},\ }\href
  {https://doi.org/10.1364/OE.387612} {\bibfield  {journal} {\bibinfo
  {journal} {Opt. Express}\ }\textbf {\bibinfo {volume} {28}},\ \bibinfo
  {pages} {11666} (\bibinfo {year} {2020})}\BibitemShut {NoStop}%
\bibitem [{\citenamefont {Yu}\ \emph {et~al.}(2021)\citenamefont {Yu},
  \citenamefont {Wang}, \citenamefont {Gao}, \citenamefont {Li}, \citenamefont
  {Zhao},\ and\ \citenamefont {Yao}}]{HadRussDoll}%
  \BibitemOpen
  \bibfield  {author} {\bibinfo {author} {\bibfnamefont {Z.}~\bibnamefont
  {Yu}}, \bibinfo {author} {\bibfnamefont {X.-Q.}\ \bibnamefont {Wang}},
  \bibinfo {author} {\bibfnamefont {C.}~\bibnamefont {Gao}}, \bibinfo {author}
  {\bibfnamefont {Z.}~\bibnamefont {Li}}, \bibinfo {author} {\bibfnamefont
  {H.}~\bibnamefont {Zhao}},\ and\ \bibinfo {author} {\bibfnamefont
  {Z.}~\bibnamefont {Yao}},\ }\bibfield  {title} {\bibinfo {title}
  {Differential hadamard ghost imaging via single-round detection},\ }\href
  {https://doi.org/10.1364/OE.441501} {\bibfield  {journal} {\bibinfo
  {journal} {Opt. Express}\ }\textbf {\bibinfo {volume} {29}},\ \bibinfo
  {pages} {41457} (\bibinfo {year} {2021})}\BibitemShut {NoStop}%
\bibitem [{\citenamefont {Yu}\ and\ \citenamefont {Liu}(2019)}]{HadOrigami}%
  \BibitemOpen
  \bibfield  {author} {\bibinfo {author} {\bibfnamefont {W.-K.}\ \bibnamefont
  {Yu}}\ and\ \bibinfo {author} {\bibfnamefont {Y.-M.}\ \bibnamefont {Liu}},\
  }\bibfield  {title} {\bibinfo {title} {Single-pixel imaging with origami
  pattern construction},\ }\href {https://doi.org/10.3390/s19235135} {\bibfield
   {journal} {\bibinfo  {journal} {Sensors}\ }\textbf {\bibinfo {volume}
  {19}},\ \bibinfo {pages} {5135} (\bibinfo {year} {2019})}\BibitemShut
  {NoStop}%
\bibitem [{\citenamefont {Yu}(2019)}]{CakeCutting}%
  \BibitemOpen
  \bibfield  {author} {\bibinfo {author} {\bibfnamefont {W.-K.}\ \bibnamefont
  {Yu}},\ }\bibfield  {title} {\bibinfo {title} {Super sub-nyquist single-pixel
  imaging by means of cake-cutting hadamard basis sort},\ }\href
  {https://doi.org/10.3390/s19194122} {\bibfield  {journal} {\bibinfo
  {journal} {Sensors}\ }\textbf {\bibinfo {volume} {19}},\ \bibinfo {pages}
  {4122} (\bibinfo {year} {2019})}\BibitemShut {NoStop}%
\bibitem [{\citenamefont {Ota}(2018)}]{ota}%
  \BibitemOpen
  \bibfield  {author} {\bibinfo {author} {\bibfnamefont {Y.}~\bibnamefont
  {Ota}, \bibfnamefont {K~;~Hayasaki}},\ }\bibfield  {title} {\bibinfo {title}
  {Complex amplitude single-pixel imaging},\ }\href@noop {} {\bibfield
  {journal} {\bibinfo  {journal} {Opt. Lett.}\ }\textbf {\bibinfo {volume}
  {43}},\ \bibinfo {pages} {3682} (\bibinfo {year} {2018})}\BibitemShut
  {NoStop}%
\bibitem [{\citenamefont {Skarsoulis}\ \emph {et~al.}(2021)\citenamefont
  {Skarsoulis}, \citenamefont {Kakkava},\ and\ \citenamefont
  {Psaltis}}]{Skarsoulis:21}%
  \BibitemOpen
  \bibfield  {author} {\bibinfo {author} {\bibfnamefont {K.}~\bibnamefont
  {Skarsoulis}}, \bibinfo {author} {\bibfnamefont {E.}~\bibnamefont
  {Kakkava}},\ and\ \bibinfo {author} {\bibfnamefont {D.}~\bibnamefont
  {Psaltis}},\ }\bibfield  {title} {\bibinfo {title} {Predicting optical
  transmission through complex scattering media from reflection patterns with
  deep neural networks},\ }\href
  {https://doi.org/https://doi.org/10.1016/j.optcom.2021.126968} {\bibfield
  {journal} {\bibinfo  {journal} {Optics Communications}\ }\textbf {\bibinfo
  {volume} {492}},\ \bibinfo {pages} {126968} (\bibinfo {year}
  {2021})}\BibitemShut {NoStop}%
\bibitem [{\citenamefont {Kanngiesser}\ \emph
  {et~al.}(2019{\natexlab{a}})\citenamefont {Kanngiesser}, \citenamefont
  {Rahlves},\ and\ \citenamefont {Roth}}]{Kanngiesser:19}%
  \BibitemOpen
  \bibfield  {author} {\bibinfo {author} {\bibfnamefont {J.}~\bibnamefont
  {Kanngiesser}}, \bibinfo {author} {\bibfnamefont {M.}~\bibnamefont
  {Rahlves}},\ and\ \bibinfo {author} {\bibfnamefont {B.}~\bibnamefont
  {Roth}},\ }\bibfield  {title} {\bibinfo {title} {Iterative wavefront
  correction for complex spectral domain optical coherence tomography},\ }\href
  {https://doi.org/10.1364/OL.44.001347} {\bibfield  {journal} {\bibinfo
  {journal} {Opt. Lett.}\ }\textbf {\bibinfo {volume} {44}},\ \bibinfo {pages}
  {1347} (\bibinfo {year} {2019}{\natexlab{a}})}\BibitemShut {NoStop}%
\bibitem [{\citenamefont {Kanngiesser}\ \emph
  {et~al.}(2019{\natexlab{b}})\citenamefont {Kanngiesser}, \citenamefont
  {Rahlves},\ and\ \citenamefont {Roth}}]{Kanngiesser:Nature}%
  \BibitemOpen
  \bibfield  {author} {\bibinfo {author} {\bibfnamefont {J.}~\bibnamefont
  {Kanngiesser}}, \bibinfo {author} {\bibfnamefont {M.}~\bibnamefont
  {Rahlves}},\ and\ \bibinfo {author} {\bibfnamefont {B.}~\bibnamefont
  {Roth}},\ }\bibfield  {title} {\bibinfo {title} {Double interferometer design
  for independent wavefront manipulation in spectral domain optical coherence
  tomography},\ }\href
  {https://doi.org/https://doi.org/10.1038/s41598-019-50996-2} {\bibfield
  {journal} {\bibinfo  {journal} {Scientific Reports}\ }\textbf {\bibinfo
  {volume} {9}},\ \bibinfo {pages} {14651} (\bibinfo {year}
  {2019}{\natexlab{b}})}\BibitemShut {NoStop}%
\bibitem [{\citenamefont {Boniface}\ \emph {et~al.}(2019)\citenamefont
  {Boniface}, \citenamefont {Blochet}, \citenamefont {Dong},\ and\
  \citenamefont {Gigan}}]{Boniface:19}%
  \BibitemOpen
  \bibfield  {author} {\bibinfo {author} {\bibfnamefont {A.}~\bibnamefont
  {Boniface}}, \bibinfo {author} {\bibfnamefont {B.}~\bibnamefont {Blochet}},
  \bibinfo {author} {\bibfnamefont {J.}~\bibnamefont {Dong}},\ and\ \bibinfo
  {author} {\bibfnamefont {S.}~\bibnamefont {Gigan}},\ }\bibfield  {title}
  {\bibinfo {title} {Noninvasive light focusing in scattering media using
  speckle variance optimization},\ }\href
  {https://doi.org/10.1364/OPTICA.6.001381} {\bibfield  {journal} {\bibinfo
  {journal} {Optica}\ }\textbf {\bibinfo {volume} {6}},\ \bibinfo {pages}
  {1381} (\bibinfo {year} {2019})}\BibitemShut {NoStop}%
\bibitem [{\citenamefont {Vellekoop}\ and\ \citenamefont
  {Mosk}(2007)}]{Vellekoop:07}%
  \BibitemOpen
  \bibfield  {author} {\bibinfo {author} {\bibfnamefont {I.~M.}\ \bibnamefont
  {Vellekoop}}\ and\ \bibinfo {author} {\bibfnamefont {A.~P.}\ \bibnamefont
  {Mosk}},\ }\bibfield  {title} {\bibinfo {title} {Focusing coherent light
  through opaque strongly scattering media},\ }\href
  {https://doi.org/10.1364/OL.32.002309} {\bibfield  {journal} {\bibinfo
  {journal} {Opt. Lett.}\ }\textbf {\bibinfo {volume} {32}},\ \bibinfo {pages}
  {2309} (\bibinfo {year} {2007})}\BibitemShut {NoStop}%
\bibitem [{\citenamefont {Popoff}\ \emph {et~al.}(2010)\citenamefont {Popoff},
  \citenamefont {Lerosey}, \citenamefont {Carminati}, \citenamefont {Fink},
  \citenamefont {Boccara},\ and\ \citenamefont {Gigan}}]{T_Mat1}%
  \BibitemOpen
  \bibfield  {author} {\bibinfo {author} {\bibfnamefont {S.~M.}\ \bibnamefont
  {Popoff}}, \bibinfo {author} {\bibfnamefont {G.}~\bibnamefont {Lerosey}},
  \bibinfo {author} {\bibfnamefont {R.}~\bibnamefont {Carminati}}, \bibinfo
  {author} {\bibfnamefont {M.}~\bibnamefont {Fink}}, \bibinfo {author}
  {\bibfnamefont {A.~C.}\ \bibnamefont {Boccara}},\ and\ \bibinfo {author}
  {\bibfnamefont {S.}~\bibnamefont {Gigan}},\ }\bibfield  {title} {\bibinfo
  {title} {Measuring the transmission matrix in optics: An approach to the
  study and control of light propagation in disordered media},\ }\href
  {https://doi.org/10.1103/PhysRevLett.104.100601} {\bibfield  {journal}
  {\bibinfo  {journal} {Phys. Rev. Lett.}\ }\textbf {\bibinfo {volume} {104}},\
  \bibinfo {pages} {100601} (\bibinfo {year} {2010})}\BibitemShut {NoStop}%
\bibitem [{\citenamefont {Brunton}\ and\ \citenamefont
  {Kutz}(2019)}]{databook}%
  \BibitemOpen
  \bibfield  {author} {\bibinfo {author} {\bibfnamefont {S.}~\bibnamefont
  {Brunton}}\ and\ \bibinfo {author} {\bibfnamefont {J.}~\bibnamefont {Kutz}},\
  }\href@noop {} {\emph {\bibinfo {title} {Data-driven Science and Engineering:
  Machine Learning, Dynamical Systems and Control}}}\ (\bibinfo  {publisher}
  {Cambridge: Cambridge University Press},\ \bibinfo {year} {2019})\BibitemShut
  {NoStop}%
\bibitem [{\citenamefont {van~den Berg}\ and\ \citenamefont
  {Friedlander}(2008)}]{BergFriedlander:2008}%
  \BibitemOpen
  \bibfield  {author} {\bibinfo {author} {\bibfnamefont {E.}~\bibnamefont
  {van~den Berg}}\ and\ \bibinfo {author} {\bibfnamefont {M.~P.}\ \bibnamefont
  {Friedlander}},\ }\bibfield  {title} {\bibinfo {title} {Probing the pareto
  frontier for basis pursuit solutions},\ }\href
  {https://doi.org/10.1137/080714488} {\bibfield  {journal} {\bibinfo
  {journal} {SIAM Journal on Scientific Computing}\ }\textbf {\bibinfo {volume}
  {31}},\ \bibinfo {pages} {890} (\bibinfo {year} {2008})}\BibitemShut
  {NoStop}%
\bibitem [{\citenamefont {van~den Berg}\ and\ \citenamefont
  {Friedlander}(2019)}]{spgl1site}%
  \BibitemOpen
  \bibfield  {author} {\bibinfo {author} {\bibfnamefont {E.}~\bibnamefont
  {van~den Berg}}\ and\ \bibinfo {author} {\bibfnamefont {M.~P.}\ \bibnamefont
  {Friedlander}},\ }\href@noop {} {\bibinfo {title} {{SPGL1}: A solver for
  large-scale sparse reconstruction}} (\bibinfo {year} {2019}),\ \bibinfo
  {note} {https://friedlander.io/spgl1}\BibitemShut {NoStop}%
\bibitem [{\citenamefont {Ren}\ \emph {et~al.}(2019)\citenamefont {Ren},
  \citenamefont {Zhao},\ and\ \citenamefont {Lam}}]{ZernnikeCS}%
  \BibitemOpen
  \bibfield  {author} {\bibinfo {author} {\bibfnamefont {Z.}~\bibnamefont
  {Ren}}, \bibinfo {author} {\bibfnamefont {J.}~\bibnamefont {Zhao}},\ and\
  \bibinfo {author} {\bibfnamefont {E.~Y.}\ \bibnamefont {Lam}},\ }\bibfield
  {title} {\bibinfo {title} {Automatic compensation of phase aberrations in
  digital holographic microscopy based on sparse optimization},\ }\href
  {https://doi.org/10.1063/1.5115079} {\bibfield  {journal} {\bibinfo
  {journal} {APL Photonics}\ }\textbf {\bibinfo {volume} {4}},\ \bibinfo
  {pages} {110808} (\bibinfo {year} {2019})},\ \Eprint
  {https://arxiv.org/abs/https://doi.org/10.1063/1.5115079}
  {https://doi.org/10.1063/1.5115079} \BibitemShut {NoStop}%
\end{thebibliography}%
%

\end{document}